\documentclass[11p,a4]{article}

\begin{document}
\input amssym.tex

\title{Covariant fields on anti-de Sitter spacetimes}

\author{Ion I. Cot\u aescu\\ {\small \it West 
                 University of Timi\c soara,}\\
   {\small \it V. P\^ arvan Ave. 4, RO-1900 Timi\c soara, Romania}}

\date{\today}
\maketitle

\begin{abstract}
The covariant free fields of any spin on anti-de Sitter spacetimes are studied, pointing out that these transform under isometries according to covariant representations of the anti-de Sitter isometry group, induced by those of the Lorentz group.  Applying the method of ladder operators it is shown that the covariant representations with unique spin  are equivalent with discrete unitary irreducible representations of positive energy of the universal covering group of the isometry one. The action of the Casimir operators is studied finding how  the weights of these representations may depend on  the mass and spin of the covariant field.  The conclusion is that on anti-de Sitter spacetime one cannot formulate an universal mass condition as in special relativity.   

Pacs: 04.02.-q and 04.02.Jb

\end{abstract}

\newpage

\section{Introduction}

The properties of the quantum fields on flat or curved spacetimes depend on their interactions among themselves and with the gravity of the  background. The mass and spin are the fundamental properties of the free fields  that may be related to the geometric invariants, the mass depending, in addition, on the manner in which  the fields are coupled to the background gravity and the effects of renormalization.  

In special relativity,  the mass and spin of the free fields are related to the weight of the unitary irreducible representations (UIR) of the Poincar\'e group, determining the eigenvalues of its Casimir operators.  Hereby it comes out  the  well-known universal mass condition defining the squared mass as the eigenvalue of the first Casimir operator. On the other hand, the quantum fields transform according to covariant representations (CR) involving linear representations (reps.) of the Lorentz group. In the physical case of  fields with a given mass and unique spin, the CRs are equivalent with UIRs, preserving thus the unitarity of the quantum theory \cite{Wigner,WKT}.  

This mechanism might work even in general relativity where the gravitation of the curved backgrounds could give rise to new relations among the geometric invariants and the mass and spin of the free fields. Unfortunately, there are serious difficulties related to the manner in which the covariant fields are defined because of the fields with semi-integer spins which do not comply with the general relativistic covariance.  These fields must be defined only in orthogonal non-holonomic frames, transforming according to the gauge group which in the case of the physical $(1+3)$-dimensional spacetimes is just the Lorentz one \cite{SW,BD}. 

In order to avoid these difficulties we proposed a general definition of CRs as reps. of the isometry group induced by finite-dimensional reps. of the gauge group \cite{ES}. Similar CRs were constructed earlier by Nachtmann for the de Sitter (dS) isometry group \cite{Nach}, adapting the Wigner method of induced reps. \cite{Wigner} but in configuration instead of momentum rep..  We have shown that both these methods are equivalent leading to the same type of  induced CRs \cite{CNach} which offer us the opportunity of defining  the spin just as in special relativity, independent on the background gravity. The advantage of this approach is of giving the most general method of constructing covariant fields of any spin on curved backgrounds and the corresponding conserved operators generating the isometry transformations. In this framework we obtained a coherent quantum theory on curved spacetimes that allowed us to  construct the dS QED in Coulomb gauge \cite{CQED}.

In general, the CRs are different from the UIRs even if these are reps. of the same isometry group.  In special relativity, the CRs are equivalent with the UIRs whose invariants determine the mass and spin of the covariant fields. Moreover, in the case of the dS spacetime we have shown that  the CRs with unique spin are equivalent with UIRs of the principal series of the $SO(1,4)$ isometry group \cite{DT1,DT2}, pointing out that the boson and fermion fields minimally coupled  to gravity do not comply with the same mass condition \cite{CGRG}. 

Here we would like to extend this study to the anti-de Sitter (AdS) spacetime where we know that the energy spectra of the Klein-Gordon \cite{AdS4,Cq1} and Dirac \cite{Cq2,Cq3} free fields are discrete, corresponding to discrete UIRs with positive energy of the $SO(2,3)$ isometry group \cite{SO231,SO232,SO233}. Our principal goal is to generalize these results showing that the induced CRs with unique spin are equivalent with this type of  UIRs  and study how the weights of these reps. depend on  mass and spin. 

For this pourpose we chose the algebraic method of ladder operators exploiting the  properties of the generators of our induced CRs which are the principal conserved observables of the quantum theory. In this manner,  we demonstrate the CR-UIR equivalence finding how the weight of UIRs may depend on mass and spin. The conclusion is that, just as in the dS case \cite{CGRG}, on AdS spacetimes we cannot postulate an universal mass condition since the bosons and fermions have different behaviors.  

The results presented here complete the global image about the theory of free fiels of any spin on the $(1+3)$-dimensional spacetimes with maximal symmetry, i. e. Minkowski, dS and AdS ones \cite{SW}. Moreover, these results can be generalized easily to the  AdS spacetimes of any dimensions involved in the  AdS/CFT-correspondence of the string theory \cite{Mal, AdSCFT}.     

We start in the second section presenting our general theory of induced CRs and their generators,  discussing briefly the methods proposed by Weinberg \cite{SWspin} and Fronsdal \cite{Fr1,Fr2,Fr3} for constructing covariant fields with unique spin. The next section is devoted to the $(1+3)$-dimensional AdS spacetime and its isometry group whose UIRs and their invariants  \cite{SO231,SO232,SO233} are revisited giving only the technical details we need for our investigation.  The original results are presented in section 4 where we demonstrate the equivalence among CRs with unique spin and UIRs. Here we give the general form of the massive covariant fields and we express the AdS invariants in terms of effective mass and spin discussing the problem of the universality of the mass condition.  In the last section we present other concluding remarks.    
      
\section{Covariant fields}

The covariant fields with integer spins on pseudo-Riemannian spacetimes   transform under isometries, according to the general relativistic covariance, as vectors or tensors of any rank defined in  holonomic frames \cite{Wald}. For the fields with  half integer spin we need to consider simultaneously both the holonomic and local non-holonomic frames which will form the fixed framework of our approach \cite{ES}. 

\subsection{Induced CRs}

The holonomic  frames are local charts of coordinates $x^{\mu}$, labeled by natural indices, $\mu, \nu,...=0,1,2,3$ while the non-holonomic frames and the corresponding dual coframes are defined by the tetrad fields $e_{\hat\mu}$  and  $\hat e^{\hat\mu}$ respectively. These  are labeled by local indices, $\hat\mu, \hat\nu,...=0,1,2,3$ and satisfy the usual duality, $\hat e^{\hat\mu}_{\alpha}\,
e_{\hat\nu}^{\alpha}=\delta^{\hat\mu}_{\hat\nu}$, $ \hat
e^{\hat\mu}_{\alpha}\, e_{\hat\mu}^{\beta}=\delta^{\beta}_{\alpha}$,
and orthonormalization,  $e_{\hat\mu}\cdot e_{\hat\nu}=\eta_{\hat\mu
\hat\nu}$, $\hat e^{\hat\mu}\cdot \hat e^{\hat\nu}=\eta^{\hat\mu
\hat\nu}$,  conditions. With their help one defines the local derivatives $\hat\partial_{\hat\alpha}=e_{\hat\alpha}^{\mu}\partial_{\mu}$ and the 1-forms $\tilde\omega^{\hat\alpha}=\hat e_{\mu}^{\hat\alpha}dx^{\mu}$. The metric tensor $g_{\mu\nu}=\eta_{\hat\alpha\hat\beta}\hat e^{\hat\alpha}_{\mu}\hat
e^{\hat\beta}_{\nu}$ raises or lowers the natural indices while for
the local indices  we have to use the flat metric $\eta$.  

The metric $\eta$ remains invariant under the transformations of the
group $O(1,3)$ which includes as a subgroup the Lorentz group,
$L_{+}^{\uparrow}$, whose universal covering group is the group $SL(2,\Bbb
C)$. In the usual covariant parametrization, with the real parameters, $\omega^{\hat\alpha
\hat\beta}=-\omega^{\hat\beta\hat\alpha}$, the transformations
$A(\omega)=\exp(-\frac{i}{2}\omega^{\hat\alpha\hat\beta}
S_{\hat\alpha\hat\beta}) \in SL(2,\Bbb C)$ depend on the covariant
basis-generators of the $sl(2,\Bbb C)$ Lie algebra, $S_{\hat\alpha
\hat\beta}$, which are the principal spin operators generating all
the spin terms of other operators. This parametrization offers us, in
addition, the advantage of a simple expansion of the matrix elements
in local bases,
$\Lambda^{\hat\mu\,\cdot}_{\cdot\,\hat\nu}[A(\omega)]=
\delta^{\hat\mu}_{\hat\nu}
+\omega^{\hat\mu\,\cdot}_{\cdot\,\hat\nu}+\cdots$, of the
transformations $\Lambda[A(\omega)]\in L_{+}^{\uparrow}$ associated
to $A(\omega)$ through the canonical homomorphism \cite{WKT}. When  $(M,g)$ is assumed to be orientable and time-orientable we may consider the Lorentz group, $L^{\uparrow}_{+}$,  as the gauge group of the Minkowski metric $\eta$ \cite{Wald}. 

Under such circumstances, we consider the covariant fields $\psi_{(\rho)}:\, M\to {\cal V}_{(\rho)}$,  locally defined over $(M,g)$ with values in the vector spaces ${\cal V}_{(\rho)}$ carrying  finite-dimensional reps. $\rho$ of the $SL(2,{\Bbb C})$ group which, in general, are reducible.  They determine the form of the covariant derivatives of the field $\psi_{(\rho)}$ in local frames,
\begin{equation}\label{der}
D_{\hat\alpha}^{(\rho)}= e_{\hat\alpha}^{\mu}D_{\mu}^{(\rho)}=\hat\partial_{\hat\alpha}
+\frac{i}{2}\, \rho(S^{\hat\beta\, \cdot} _{\cdot\,
\hat\gamma})\,\hat\Gamma^{\hat\gamma}_{\hat\alpha \hat\beta}\,.
\end{equation}
which depend on the connection coefficients  in local frames
\begin{equation}
\hat\Gamma^{\hat\sigma}_{\hat\mu \hat\nu}=e_{\hat\mu}^{\alpha}
e_{\hat\nu}^{\beta}(\hat e_{\gamma}^{\hat\sigma}
\Gamma^{\gamma}_{\alpha \beta} -\hat e^{\hat\sigma}_{\beta,
\alpha})\,, 
\end{equation}
where $\Gamma^{\gamma}_{\alpha \beta}$ denote the Christoffel symbols. These covariant derivative  assure the covariance of the whole theory under the (point-dependent) tetrad-gauge transformations, 
\begin{eqnarray}
\tilde\omega & \to &{\tilde\omega}'= \Lambda[A]\tilde\omega\,,\\
\psi_{(\rho)}&\to & \psi_{(\rho)}'=  \rho(A)\psi_{(\rho)}\,,\label{Gauge}
\end{eqnarray}
produced by the sections $A\in SL(2,{\Bbb C})$ of the spin fiber bundle \cite{Wald}.

The isometries, $x\to x'=\phi_{\xi}(x)$, depend on the parameters $\xi^a$ ($a,b,...=1,2...N$)  of the isometry group  $I(M)$ of the manifold $(M,g)$. We have shown that each isometry  must combined with a gauge transformation $A_{\xi}\in SL(2,\Bbb C)$ in order to restore the initial relative position between the natural and local frames. We deduced that this transformation is defined as \cite{ES}
\begin{equation}\label{Axx}
\Lambda^{\hat\alpha\,\cdot}_{\cdot\,\hat\beta}[A_{\xi}(x)]= \hat
e_{\mu}^{\hat\alpha}[\phi_{\xi}(x)]\frac{\partial
\phi^{\mu}_{\xi}(x)} {\partial x^{\nu}}\,e^{\nu}_{\hat\beta}(x)\,,
\end{equation}
with the supplementary condition $A_{\xi=0}(x)=1\in SL(2,\Bbb C)$.  Then, the combined 
transformations $(A_{\xi},\phi_{\xi})$ preserve the gauge,  
\begin{equation}
(A_{\xi},\phi_{\xi}):\quad
\begin{array}{rlrcl}
e(x)&\to&e'(x')&=&e[\phi_{\xi}(x)]\,,\\
\hat e(x)&\to&\hat e'(x')&=&\hat e[\phi_{\xi}(x)]\,,
\end{array}
\qquad
\end{equation}
transforming the covariant fields according to the rule
\begin{equation}
(A_{\xi},\phi_{\xi}):\quad \psi_{(\rho)}(x)\to\psi_{(\rho)}'(x')=\rho[A_{\xi}(x)]\psi_{(\rho)}(x)\,.\label{es}
\end{equation}
which defines the operator-valued CR  $T^{(\rho)} \,:\, (A_{\xi},\phi_{\xi})\to T_{\xi}^{(\rho)}$  whose operators act as
\begin{equation}\label{Tx}
(T_{\xi}^{(\rho)}\psi_{(\rho)})[\phi_{\xi}(x)]=\rho[A_{\xi}(x)]\psi_{(\rho)}(x)\,,
\end{equation}
We specify that the pairs $(A_{\xi},\phi_{\xi})$ constitute a well-defined Lie group we called the external symmetry group of $(M,g)$, denoted by $S(M)$, pointing out that this is isomorphic with the universal covering group of the isometry group $I(M)$ \cite{ES}. 

Thus, we constructed the  CRs of the group $S(M)$, induced by the finite-dimensional reps. $\rho$ of the group $SL(2,{\Bbb C})$, that may be used for the fields with semi-integer spins.  In the case of the fields with integer spins these CRs are equivalent with the usual vector and tensor reps. of general relativity \cite{ES}.  

For small values of $\xi^{a}$, we can expand the isometres,  $x\to x'=\phi_{\xi}(x)=x+\xi^a k_a(x) +...$, in terms of  the Killing vectors, $k_a$, associated with this parametrization. Then, the parameters of the transformations $A_{\xi}(x)\equiv A[\omega_{\xi}(x)]$ can be expanded in their turn as $\omega^{\hat\alpha\hat\beta}_{\xi}(x)=\xi^{a}\Omega^{\hat\alpha\hat\beta}_{a}(x)+\cdots$, laying out of the new functions
\begin{equation}\label{Om}
\Omega^{\hat\alpha\hat\beta}_{a}\equiv {\frac{\partial
\omega^{\hat\alpha\hat\beta}_{\xi}} {\partial\xi^a}}_{|\xi=0}
=\left( \hat e^{\hat\alpha}_{\mu}\,k_{a,\nu}^{\mu} +\hat
e^{\hat\alpha}_{\nu,\mu}
k_{a}^{\mu}\right)e^{\nu}_{\hat\lambda}\eta^{\hat\lambda\hat\beta}
\end{equation}
which depend only on the Killing vectors and tetrades \cite{ES}. With their help we may write down the generators of the induced CRs,
\begin{equation}\label{Sx}
X_{a}^{(\rho)}=i{\partial_{\xi^a} T_{\xi}^{(\rho)}}_{|\xi=0}=-i
k_a^{\mu}\partial_{\mu} +\frac{1}{2}\,\Omega^{\hat\alpha\hat\beta}_{a}
\rho(S_{\hat\alpha\hat\beta})\,.
\end{equation}
These generators  satisfy the commutation rules $[X_{a}^{(\rho)},
X_{b}^{(\rho)}]=ic_{abc}X_{c}^{(\rho)}$ determined by the structure
constants, $c_{abc}$, of the algebras $s(M)\sim i(M)$. In other
words, the operators (\ref{Sx}) are the basis-generators of a CR of
the $s(M)$ algebra {induced} by the rep. $\rho$ of the
$sl(2,{\Bbb C})$ algebra.

We note that the generators (\ref{Sx}) are proportional with the  Kosmann's Lie derivatives \cite{Kos} associated to the Killing vectors $k_a$. They can be put in  covariant form either in non-holonomic frames \cite{ES},
\begin{equation}
X^{(\rho)}_{a}=-ik^{\mu}_{a}D_{\mu}^{(\rho)}+\frac{1}{2}\,
k_{a\, \mu;\nu}\,e^{\mu}_{\hat\alpha}\,e^{\nu}_{\hat\beta}\,
\rho(S^{\hat\alpha\hat\beta})
\end{equation}
or even in holonomic ones \cite{EPL}, generalizing thus the formula given by Carter and McLenaghan for the Dirac field \cite{CML}.

\subsection{Covariant fields with unique spin}

The above definition of the CRs transforming the covariant fields is general including all the particular cases studied so far. Thus the covariant fields with  integer spin which are independent on the local frames are just the vectors and tensors of any rank transforming covariantly under isometries, as we have shown in Ref. \cite{ES}. Therefore, the CRs are useful especially in theories involving covariant fields  with half integer spin, depending on the choice of the orthogonal local frames.  On the other hand, our approach gives the general rule (\ref{Sx}) of deriving the isometry generators which represent the principal observables of the quantum theory.  These operators are conserved in the sense that they commute with the operators of the field equations resulted from an invariant Lagrangian theory.

In general, the Lagrangian densities may be constructed with the help of some  positive defined quadratic forms which must remain  invariant under the transformations (\ref{es})  induced by the reps. $\rho$. Since all these  are non-unitary,  we need to use reducible reps. and the (generalized) Dirac conjugation, $\overline\psi_{(\rho)}=\psi^+_{\rho} \gamma_{(\rho)}$, where the matrix $\gamma_{(\rho)}=\gamma^+_{(\rho)}=\gamma^{-1}_{(\rho)}$ must be chosen  such that  
\begin{equation}
\overline{\rho (A)}=\gamma_{(\rho)}\rho(A)^+\gamma_{(\rho)}=\rho(A^{-1})
\end{equation}
Then the generators of the rep. $\rho$ are self-adjoint with respect to the Dirac conjugation, $\overline{\rho(S)}=\rho(S)$, and  the quadratic forms $\overline\psi_{(\rho)}\cdots \psi_{(\rho)}$ are invariant under the  transformations (\ref{es}). The Dirac conjugation can be defined  either for self-adjoint irreps. $(j,j)$ or for the {symmetric} reps., $\rho=...(j_1,j_2)\oplus (j_2,j_1)...$, which are direct sums formed excursively  by pairs of adjoint irreps., as we briefly argue in the Appendix. In this manner the spin content of the theory, denoted by ${\Bbb S}(\rho)$, is increasing since each irrep.  $ (j_1,j_2)$ brings the subspeces ${\cal V}_s$  of the UIRs of the group $SU(2)$ with spins  $s=j_1+j_2, j_1+j_2-1,...|j_1-j_2|$  \cite{WKT}. Then the carrier space of the rep. $\rho$ can be decomposed as
\begin{equation}
{\cal V}_{(\rho)}=\sum_{s\in {\Bbb S}(\rho)}\oplus {\cal V}_{\sigma}\,.
\end{equation}
We remind the reader that the irreducible reps. (irreps.) with unique spin $s$ are only $(s,0)$ and $(0,s)$. 

Under such circumstances it is difficult to construct covariant fields with {\em unique} spin $s$, eliminating the unwanted components.  The simplest method was proposed by Weinberg \cite{SWspin} which assumed that a covariant field of spin $s$ transforms according to the CR induced by the rep. $\rho_s=(s,0)\oplus (0,s)$.  In Minkowski spacetimes these fields satisfy field equations with derivatives up to the order $2s$ that can be rewritten in curved spacetimes by replacing the usual derivatives with covariant ones. The typical example is the Dirac field  with $s=\frac{1}{2}$ and $\rho_D=(\frac{1}{2},0)\oplus(0,\frac{1}{2})$ whose AdS quantum modes in Cartesian gauge we derived long time ago \cite{Cq2}.  However,  for $s>1$ the field equations of the order $2s>2$ may have unwanted solutions of the dipole-ghost type that do not have a physical meaning at the level of the theory of free fields and cannot be eliminated by imposing supplemental symmetries  \cite{Cghost}.   Thus the Weinberg method works successfully  for the usual Dirac and Proca or Maxwell fields while for the higher spin fields the problem of eliminating the ghosts remains open.

An alternative method that works for covariant fields with spins $s>1$   was  proposed by Fronsdal in Minkowski spacetime \cite{Fr1} generalizing the Pauli-Fiertz theory of the fields with spin $\frac{3}{2}$ \cite{PaFi}. The idea is to exploit self-adjoint irreps. $(\frac{s}{2},\frac{s}{2})$ for the fields of integer spin $s$ eliminating the lower spin components, with spins $s-1,s-2,...0$, by using special field equations and trace conditions. The advantage is that this theory can be formulated in terms of symmetric tensors of the ranks  $s$ for which the field equations and the trace conditions can be written in covariant form that can be  generalized to curved manifolds. Thus one obtains good field equations of the second order whose coefficients depend on $s$. Unfortunately, this method works only for massless fields but it is very useful and popular solving the problem of the graviton of spin $s=2$ without resorting to equations with higher order derivatives. A similar method can be applied in the case of the covariant massless fields with half integer spin $s$, starting with the rep. $\rho=\rho_D \otimes (\frac{s}{2}-\frac{1}{4},\frac{s}{2}-\frac{1}{4})$ of the symmetric Rarita-Schwinger spinor-tensors of the rank $s-\frac{1}{2}$ \cite{Fr2,Fr3} that represent the natural generalization of the Pauli-Fiertz spinor-vector.  

Hereby we may conclude that the general problem of constructing Lagrangian theories of massive covariant fields with unique spin and field equations of at most second order derivatives is not yet solved. Nevertheless, we can study the Weinberg covariant fields with unique spin from the mathematical point of view focusing on the CR-UIR equivalence as the first step to a complete quantum theory of covariant fields. We have seen that,  in special relativity and on dS spacetime \cite{CGRG},  the CRs with unique spin are equivalent with several UIRs whose invariants get a physical meaning. In what follows we would like to study  the same problem  in the case of  CAdS spacetimes for understanding how the mass and spin of the covariant fields may be defined in this geometry.

\section{Anti-de Sitter isometry group}

Let us focus now on the CAdS spacetime $(M,g)$  defined as the universal covering space of the $(1+3)$-dimensional AdS spacetime. This is a vacuum solution of the Einstein equations with $\Lambda<0$ and negative constant curvature, representing a hyperboloid of radius $R=\frac{1}{\omega} = \sqrt{-\frac{3}{\Lambda}}$  embedded  in the five-dimensional pseudo-Euclidean spacetime $(M^5,\eta^5)$ of metric $\eta^5={\rm diag}(1,1,-1,-1,-1)$ where we consider the Cartesian coordinates $z^A$  ($A,\,B,...= -1,0,1,2,3$). 

\subsection{CADS spacetime}

On CAdS spacetimes we may introduce local coordinates, $x^{\mu}$ ($\alpha,...\mu,\nu...=0,1,2,3$)  giving the functions $z^A(x)$ which solve the hyperboloid equation,
\begin{equation}\label{hip}
\eta^5_{AB}z^A(x) z^B(x)=\frac{1}{\omega^2}\,.
\end{equation}
Here we consider only the local  chart $\{t,\bf{x}\}$ with Cartesian spaces coordinates $x^i$ ($i,j,k,...=1,2,3$) and $t\in {\Bbb R}^+$, defined by the functions
\begin{eqnarray}
z^{-1}(x)&=&\frac{1}{\omega}\chi(r)\cos(\omega t)\,,\\
z^0(x)&=&\frac{1}{\omega}\chi(r)\sin(\omega t)\,,\\
z^i(x)&=&x^i \,, \label{Zx}
\end{eqnarray}
where we denote $r=|\bf{x}|$ and $\chi(r)=\sqrt{1+\omega^2 r^2}$.  Hereby we obtain the line element,  
\begin{eqnarray}
ds^{2}&=&\eta^5_{AB}dz^A(x_c)dz^B(x_c)\nonumber\\
&=&\chi(r)^2 dt^{2}-\left[\delta_{ij}-\frac{\omega^2 x^{i}x^{j}}{\chi(r)^2}\right]dx^{i}dx^{j}\,,\label{line1}
\end{eqnarray}
laying out an obvious symmetry under space rotations and time translations called here central symmety. In order to keep this global symmetry, we chose the Cartesian tetrad gauge in which  the non-vanishing tetrad components read \cite{ES}
\begin{eqnarray}
&\hat e^{0}_0=\chi\,,&\quad  \hat e^{i}_j=\delta^i_j-\frac{x^i x^j}{r^2}\left(1-\frac{1}{\chi}\right)\,,\label{tet1}\\
& e^{0}_0=\frac{1}{\chi}\,,&\quad   e^{i}_j=\delta^i_j-\frac{x^i x^j}{r^2}\left(1-{\chi}\right)\,.\label{tet2}
\end{eqnarray}

Notice that in the associated central chart $\{t,r,\theta,\phi\}$ with spherical coordinates, canonically related to the Cartesian ones, $\bf{x}\to (r,\theta,\phi)$, we find the familiar   line element 
\begin{equation}\label{line2}
ds^2=\chi(r)^2 dt^2-\frac{dr^2}{\chi(r)^2}-r^2(d\theta^2+\sin ^2 \theta\, d\phi^2)\,.
\end{equation}

The symmetries of these manifolds are  given by the transformations ${\frak g}\in SO(2,3)$ which leave invariant the metric $\eta^5$ of the embedding manifold $(M^5,\eta^5)$ and implicitly Eq. (\ref{hip}). For these transformations we adopt the parametrization
\begin{equation}
{\frak g}(\xi)=\exp\left(-\frac{i}{2}\,\xi^{AB}{\frak S}_{AB}\right)\in SO(2,3) 
\end{equation}
with skew-symmetric parameters, $\xi^{AB}=-\xi^{BA}$,  and the covariant generators ${\frak S}_{AB}$ of the fundamental rep. of the $so(2,3)$ algebra carried by $M^5$ that have the matrix elements, 
\begin{equation}
({\frak S}_{AB})^{C\,\cdot}_{\cdot\,D}=i\left(\delta^C_A\, \eta_{BD}^5
-\delta^C_B\, \eta_{AD}^5\right)\,.
\end{equation}
In general, each transformation ${\frak g}\in SO(2,3)$ generates an isometry changing the coordinates of a local chart $\{x\}$,  defined by the functions $z=z(x)$,  according to the rule $x\to x'=\phi_{\frak g}(x)$ where the functions $\phi_{\frak g}$ are derived from the system of equations $z[\phi_{\frak g}(x)]={\frak g}z(x)$. 

Thus we understand that the $(1+3)$-dimensional CAdS spacetime has the isometry group  $I(M)=SO(2,3)$ whose universal covering group is isomorphic with the group of combined transformations, $S(M)\simeq{\rm Spin}(2,3)$. The CRs of this group are induced by the finite-dimensional reps. of the group $SL(2,{\Bbb C})$.

\subsection{UIRs of the group Spin$(2,3)$}

Our principal objective here is to study the CR-UIR equivalence of the reps. of the group  ${\rm Spin}(2,3)$ such that we must review first the UIRs of this group which are well studied from long time \cite{SO231,SO232,SO233}.  

Let us denote by $\upsilon$ the UIR of the spin$(2,3)$ Lie algebra whose  basis-generators
\begin{equation}
X_{(AB)}=\upsilon({\frak S}_{AB})=\upsilon({\frak S}_{AB})^{\dagger}
\end{equation}
are Hermitian operators with respect to  the scalar product of the carrier Hilbert space ${\cal H}$. These operators satisfy the canonical commutation rules 
\begin{eqnarray}
\left[ X_{(AB)},X_{(CD)}\right]&=&\eta_{AC}X_{(BD)}-\eta_{AD}X_{(BC)}\nonumber\\
&+&\eta_{BD}X_{(AC)}-\eta_{BC}X_{(AD)}\,.
\end{eqnarray}
The invariants of this rep. are the Casimir operators 
\begin{equation}
{\cal C}_1=\omega^2
\frac{1}{2}\,X_{(AB)}X^{(AB)}\,,\label{C1}
\end{equation}
and 
\begin{equation}\label{Q2}
{\cal C}_2=-\eta^5_{AB}W^{ A}W^{B}\,,
\end{equation}
where the operators
\begin{equation}\label{WW}
W^{A}=\frac{1}{8}\,\omega\, \varepsilon^{ABCDE}
X_{(BC)}X_{(DE)}\,,
\end{equation}
play the same role as the components of the Pauli-Lubanski four-vector of the Poincar\' e algebra.  

The generators with an useful physical meaning are the energy operator $H=\omega X_{(0,-1)}$, the total angular momentum
\begin{equation}\label{Jia}
J_i =  \frac{1}{2}\,\varepsilon_{ijk}
X_{(j,k)}\,,
\end{equation}
and the ladder operators $A_i= \omega(X_{(-1,i)}-i X_{(0,i)})$. All these operators
form the canonical basis $\{H, J_i,A_i,{A_i}^{\dagger} \}$ of the $so(2,3)$ algebra, having the  commutators
\begin{equation}
[H,J_i]=0\,, \quad
[J_i, A_j]=i\varepsilon_{ijk}A_k
\end{equation}
and
\begin{eqnarray}
&&\left[ H, A_i \right]=-\omega A_i\,,\label{HA1}\\
&& \left[H, {A_i}^{\dagger}
\right]=\omega{A_i}^{\dagger}\,,\label{HA2}\\
&&\left[A_i, A_j \right]=[{A_i}^{\dagger},{A_j}^{\dagger}]=0\,,\label{AAAA}\\
&&\left[A_i, {A_j}^{\dagger}\right]=2  \omega \delta_{ij} H - 2 i
\omega^2 \varepsilon_{ijk} J_k \,.\label{AAHJ}
\end{eqnarray}
Moreover, the Casimir operators can be rewritten as
\begin{eqnarray}
{\cal C}_1&=&H^2-3 \omega H-{\bf{A^{\dagger}}}\cdot{\bf{A}}+\omega^2{\bf{J}\,}^2\,,\label{C1a}\\
{\cal C}_2&=&{\bf{J}\,}^2\left({\cal C}_1-\omega^2 {\bf{J}\,}^2+2\omega^2\right) +{\bf F}(X)\cdot\bf{ A}\,,\label{C2a}
\end{eqnarray}
where $F_i(X)$ are operators depending on the above basis-generators.

For any space dimension $i$, the set $(H, A_i, {A_i}^{\dagger})$ can be seen as a Heisenberg-type algebra generating  oscillations on this dimension \cite{SO231}.  What is new here is that these algebras are not independent each other  because of the last term of Eq. (\ref{AAHJ}). This is not an impediment for applying  the standard procedure for determining the energy spectra staring with a subspace of ground states $\Phi_0^s$ which satisfy the conditions  
\begin{equation}\label{APhi}
A_i\Phi_0^s=0\,, \quad i=1,2,3\,.
\end{equation}    
and the eigenvalues problems \cite{SO231,SO232},
\begin{eqnarray}
H\Phi_0^s &=&E_0\Phi_0^s\,,\quad E_0> 0\\
{\bf{J}\,}^2\Phi_0^s &=&s(s+1)\Phi_0^s\,, \quad s=\textstyle{0,\frac{1}{2},1,\frac{3}{2},...}\label{JJunit}\,.
\end{eqnarray}
Hereby, it results that the UIRs, denoted by $(E_0,s)$,  are completely determined by the ground energy $E_0$ and the spin $s$  giving the principal invariants, i. e. the eigenvalues of the Casimir operators \cite{SO233},
\begin{eqnarray}
c_1&=&E_0^2 -3\omega E_0+\omega^2 s(s+1)\,,\label{cc1}\\
c_2&=&s(s+1)\left(E_0^2-3\omega E_0+2\omega^2\right)\,,\label{cc2}
\end{eqnarray}  
resulted from the eigenvalues problems ${\cal C}_{1,2}\Phi_0^s=c_{1,2}\Phi_0^s$. 

The operators of the UIR $(E_0,s)$ act in a Hilbert space whose basis can be introduced  observing that the new states 
\begin{eqnarray}
\Phi^s_{n_1n_2n_3}&=&N_{n_1n_2n_3}\nonumber\\
&\times& \left( {A_1}^{\dagger}\right)^{n_1}
\left( {A_2}^{\dagger}\right)^{n_2}\left( {A_3}^{\dagger}\right)^{n_3}
 \Phi_0^s\,,\label{genstate}
 \end{eqnarray}
where $N_{n_1n_2n_3}$ are normalization factors,  are energy eigenstates, 
\begin{equation}
H \Phi_{n_1n_2n_3}^s=\left[E_0+\omega(n_1+n_2+n_3)\right] \Phi_{n_1n_2n_3}^s\,,
\end{equation} 
as it results from  Eq. (\ref{HA2}). One obtains thus the energy spectra $E_n=E_0+ n\omega$ depending only on the principal quantum number $n=n1+n2+n3$ which means that  these spectra are deeply degenerated as in the non-relativistic case \cite{SO233}. 
 
\section{Covariant fields on CAdS spacetimes}

From our general theory it results that the covariant fields, $\psi_{(\rho)}$, on CAdS spacetimes transforms according to CRs of the Spin$(2,3)$ group induced  by finite-dimensional reps., $\rho$, of the group $SL(2,{\Bbb C})$. In what follows we denote for brevity the generators of these reps. as $S^{(\rho)}_{\hat\alpha\hat\beta}=\rho(S_{\hat\alpha\hat\beta})$.

\subsection{CRs of the Spin$(2,3)$ group}

The generators of the CRs of the group Spin$(2,3)$ depend on the Killing vectors of the CAdS spacetime that can be derived easily as \cite{CAdS1}
\begin{equation}
k_{(AB)\,\mu}=z_B\partial_{\mu} z_A -z_A\partial_{\mu} z_B\,,
\end{equation}
where $ z_A=\eta^5_{AC}z^C$.  They were calculated in  Ref. \cite{ES} according to Eq. (\ref{Sx}) and functions (\ref{Om}) with the new labels $a\to (AB)$. Thus we obtained 
the energy (or Hamiltonian) operator, 
\begin{equation}\label{Ham}
H=\omega X_{(0,-1)}^{(\rho)}=i\partial_t
\end{equation}  
and the total angular momentum,
\begin{equation}\label{Ji}
J^{(\rho)}_i \equiv  \frac{1}{2}\,\varepsilon_{ijk}
X_{(j,k)}^{(\rho)}=-i\varepsilon_{ijk}x^j\partial_k+S_i^{(\rho)}\,,
\end{equation}
where    
\begin{equation}
S^{(\rho)}_i=\frac{1}{2}\varepsilon_{ijk}S_{ij}^{(\rho)}\,. 
\end{equation}
We remained with two sets of Lorentz-type generators,
\begin{equation}
K_i^{(\rho)}=X^{(\rho)}_{(0,i)}\,, \quad N_i^{(\rho)}=X^{(\rho)}_{(-1,i)}\,,
\end{equation}
having more complicated expressions  \cite{ES} but whose algebra can be studied by using algebraic codes on computer. Notice that in the central chart we consider here the operator $H$ is the only genuine orbital operator without spin terms.

These generators form the basis $\{H,J^{(\rho)}_i,K^{(\rho)}_i,N^{(\rho)}_i\}$ of the CR induced by $\rho$ of the Lie algebra ${\rm spin}(2,3)$.  Moreover, we can introduce the canonical basis $\{H, J^{(\rho)}_i,A_i^{(\rho)},\bar{A}_i^{(\rho)} \}$ defining the ladder operators of this rep.,
\begin{eqnarray}
A_i^{(\rho)}&=& \omega(N^{(\rho)}_i-i K^{(\rho)}_i)\,,\\
\bar{A}_i^{(\rho)}&=& \omega(N^{(\rho)}_i+i K^{(\rho)}_i)\,,
\end{eqnarray}
which are no longer related through Hermitian conjugation since $\rho$ is not unitary. With these generators we can derive the Casimir operators according to Eqs. (\ref{C1a}) and (\ref{C2a}) where we have to use $\bar{A}_i^{(\rho)}$ instead of ${A_i}^{\dagger}$.

The generators of CRs have the remarkable property that in the flat limit become just the usual  generators of the CRs of the Poincar\' e group. Indeed, we observe first that the generators (\ref{Ham}) and (\ref{Ji}) are independent on $\omega$ having the same form as in the flat  case, $H=\hat H$ and $J_k^{(\rho)}=\hat J_k^{(\rho)}$. The other generators have the limits
\begin{eqnarray}
&&\lim_{\omega\to 0}(\omega N_i^{(\rho)})=\hat P^i=-i\partial_i\,, \\
&&\lim_{\omega\to 0}K_i^{(\rho)}=\hat K_i^{(\rho)} \,,
\end{eqnarray}
recovering thus the Poincare generators  $\hat H$, $\hat
P^i$, $\hat J^{(\rho)}_i$ and $\hat K^{(\rho)}_i$. Moreover, in this limit, the Casimir operators become the Poincar\'e ones
\begin{eqnarray}
&&\lim_{\omega\to 0}{\cal C}_1^{(\rho)}=\hat P^2=m^2\,,\\
&& \lim_{\omega\to 0}{\cal C}_2^{(\rho)}=\hat P^2 ({\bf{S}^{(\rho)}\,})^2=m^2 ({\bf{S}^{(\rho)}\,})^2\,,
\end{eqnarray}
suggesting that their physical meaning may be related to the mass and spin of the matter fields in a similar manner as in special relativity.

\subsection{CR-UIR equivalence}

Following  the standard procedure for determining the energy spectra  we look for the subspace ${\cal V}_0$ of the ground states $\psi_0\in {\cal V}_0\subset {\cal V}_{(\rho)}$ which satisfy
\begin{equation}\label{APhia}
A_i^{(\rho)}\psi_0=0\,, \quad i=1,2,3\,.
\end{equation}    
and
\begin{equation}\label{HPhi}
H\psi_0=E_0 \psi_0\,,~~\to~~ \psi_0=e^{-iE_0t}u_0
\end{equation}
where $E_0>0$ is the ground energy.  

The next step is to calculate the Casimir operators (\ref{C1a}) and (\ref{C2a}) but we observe that we cannot use directly Eq. (\ref{JJunit}) since now the operators $J_i^{(\rho)}$ have orbital parts whose action must be evaluated in order to find how these generators  act on the ground states $\psi_0$. This problem is complicated and cannot be solved in the general case of any rep. $\rho$. Therefore, we restrict ourselves  to the reps. with unique spin $(s,0)$ and $(0,s)$  for which we have $-S_{0i}^{(s,0)}=S_{0i}^{(0,s)}=iS_i^s$ where $S^s_i$ are the generators of the UIR of spin $s$ of the $SU(2)$ group. Under such circumstances,  our algebraic codes on computer allowed us to find the useful identity
\begin{equation}
J_i^{(s,0)}=e^{-i\omega t}\frac{1}{\chi}\,\varepsilon_{ijk}x^j A_k^{(s,0)}+\Sigma_i
\end{equation}
where $\Sigma_i$ are point-dependent matrices that in our gauge given by Eqs. (\ref{tet1}) and (\ref{tet2}) read
\begin{equation}
\Sigma_i=\frac{1}{\chi}\,e_i^l\left[S_l^s-{\omega}\varepsilon_{ljk}x^j S_{k}^s\right]\,.
\end{equation}
These matrices form a UIR of spin $s$ of the $su(2)$ algebra, since they satisfy $[\Sigma_i,\Sigma_j]=i\varepsilon_{ijk}\Sigma_k$ and  ${\bf \Sigma}^2=s(s+1) {\bf 1}_{s}$, having, in addition, the remarkable property
\begin{equation}
\left[J_i^{(s,0)}, \Sigma_i\right]=0,\quad i=1,2,3\,.
\end{equation}
Then, according to Eq. (\ref{APhia}), we my write  
\begin{equation}\label{JPhi}
J_i^{(s,0)}\psi_0=\Sigma_i \psi_0~ \to~ ({\bf J}^{(s,0)})^2\psi_0=s(s+1) \psi_0\,, 
\end{equation}
such that we can calculate the action of the Casimir operators on the ground state $\psi_0$ finding  just the eigenvalues  (\ref{cc1}) and (\ref{cc2}). Similar results can be derived for the irrep. $(0,s)$ and, therefore, for $\rho_s=(s,0)\oplus (0,s)$.

Thus we obtain our principal result showing that the CRs with ground energy $E_0$ and unique spin $s$, induced by the irreps. $(s,0)$ or $(0,s)$ of the group $SL(2,{\Bbb C})$, are equivalent with  the discrete UIRs of positive energy, $(E_0,s)$, of the group ${\rm Spin}(2,3)$ \cite{SO233}. This result can be generalized to any induced CR since any rep. $\rho$ is as a direct sum of irreps. $(j_1,j_2)=(j_1,0)\otimes (0,j_2)$ which are direct products of irreps. with unique spin \cite{WKT}. Thus we may conclude that any CR of ground energy $E_0$  induced by the rep. $\rho$ is equivalent with the reducible unitary rep.
\begin{equation}
\sum_{s\in {\Bbb S}(\rho)}\oplus (E_0,s)\,.
\end{equation}

Note that this CR-UIR equivalence established for reps. with unique spin  is similar to that we found on the dS spacetimes  where the CRs with unique spin are equivalent with UIRs from the principal series of the group ${\rm Spin}(1,4)$ \cite{DT1,DT2}, whose weights are determined by the rest energy and spin \cite{CGRG}. Moreover, the corresponding dS invariants can be obtained performing the substitution $\omega \to- i\omega$ in Eqs. (\ref{C1a}) and (\ref{C2a}), checking up again the dS-CAdS symmetry with respect to this change.

\subsection{Massive covariant fields} 

The above results allow us to write down the general form of a covariant field of spin $s$ on CAdS spacetime
\begin{equation}
\psi_{(\rho_s)}=\sum_{n,\nu} \left(U_{n,\nu}(x)a_{n,\nu}+V_{n,\nu}(x)b^*_{n,\nu}\right)\,,
\end{equation}
where $n$ is the principal quantum number while $\nu$ represent the other quantum numbers which depend on the manner in which we define the basis of the subspace with fixed $n$. More specific, by using the algebraic method, we obtain the vector-functions of positive frequencies 
\begin{equation}
U_{n_1,n_2,n_3}=N_{n_1,n_2,n_3}\left({\bar A}_1^{(\rho)}\right)^{n_1} \left({\bar A}_2^{(\rho)}\right)^{n_2}\left({\bar A}_3^{(\rho)}\right)^{n_3}\psi_0\,, 
\end{equation}
where $\psi_0: M\to {\cal V}_{(\rho_s)}$ satisfies Eqs. (\ref{APhia}),  (\ref{HPhi}) and (\ref{JPhi}). Then we separate the subspace of the vector-functions with the same $n=n_1+n_2+n_3$ where we introduce a convenient basis $\{U_{n,\nu}\}$  labeled by the quantum numbers $\nu$. The vector functions of negative frequencies may be defined as $V_{n,\nu}={\cal C}_{(\rho)}U_{n,\nu}^*$ with the help of the charge-conjugation matrix given in Appendix \cite{CdSnew}. 

The algebraic method used here is not able to give information about the mass of the covariant field since the integration constants of the Eq. (\ref{APhi}) remain arbitrary. Therefore, these constants, including the field mass,  must be determined by a specific field equation derived from a Lagrangian theory.  We remind the reader that all the conserved operators, including $A_i^{(\rho)}$ and $\bar{A}_i^{(\rho)}$, will commute with the operator of this equation such that $U_{n,\nu}$ and $V_{n,\nu}$ have to form the corresponding set of fundamental solutions. In what follows we analyze the simplest field equations looking for a general rule of defining the mass  as in special relativity where the first Casimir operator of the Poincar\'e group gives the universal mass condition $\hat P^2=m^2$.

For discussing this problem in the case of the CAdS spacetimes it is convenient to introduce the effective mass $M$  determining the ground energy as, 
\begin{equation}
E_0=M+\frac{3}{2}\,\omega\,,
\end{equation}
and bringing the energy spectra in the canonical form
\begin{equation}
E_n=M+\omega\left(n+\frac{3}{2}\right)\,,
\end{equation} 
while the invariants (\ref{cc1}) and (\ref{cc2}) of a covariant field of spin $s$ become
\begin{eqnarray}
c_1&=&M^2 -\frac{9}{4}\,\omega^2+\omega^2 s(s+1)\,,\label{cc3}\\
c_2&=&s(s+1)\left(M^2-\frac{1}{4}\omega^2\right)\,.\label{cc4}
\end{eqnarray} 

The simplest examples are the Klein-Gordon, Dirac and proca fields. For the Klein-Gordon field (with $s=0$)  minimally coupled to the CAdS gravity we find that the first Casimir operator is just the Klein-Gordon one, ${\cal C}_1={\cal E}_{KG}$, such that $c_1=m^2$.  Moreover, our preliminary calculations indicate that a similar mass condition may hold in the case of the Proca field. 

On the contrary, the operator ${\cal E}_D=i \gamma^{\hat\alpha}D^{(\rho_D)}_{\hat\alpha} $ of the Dirac equation, $({\cal E}_D -m)\psi_D=0$,  in  minimal coupling, satisfies the identities
\begin{eqnarray}
{\cal C}^{(\rho_{D})}_1&=&{\cal E}_D^2-\frac{3}{4}\,\omega^2 {\bf 1}_{\rho_{D}}\,,\\
{\cal C}^{(\rho_{D})}_2&=&\frac{3}{4}\,{\cal E}_D^2-\frac{3}{16}\omega^2 {\bf 1}_{\rho_{D}}\,.
\end{eqnarray}   
from which we deduce that now we must take $M=m$. 

A similar situation we met on the dS spacetimes where  we obtained similar expressions respecting the symmetry under the change $\omega \to -i\omega$ \cite{CGRG}.   We summarize all these results in the following table which lays out the differences between bosons and fermions on dS and CAdS spacetimes. 

\begin{center}
\begin{tabular}{lll}
CAdS&Minkowski \cite{WKT}&de Sitter \cite{CGRG}\\
{\bf Klein-Gordon}&&\\
$c_1=m^2$&$c_1=m^2$&$c_1=m^2$\\
$c_2=0$ & $c_2=0$ & $c_2=0$\\
{\bf Dirac}&&\\
$c_1=m^2-\frac{3}{2}\omega^2$&$c_1=m^2$&$c_1=m^2+\frac{3}{2}\omega^2$\\
$c_2=\frac{3}{4}\left(m^2-\frac{1}{4}\omega^2\right)$~ & $c_2=\frac{3}{4}m^2$ & $c_2=\frac{3}{4}\left(m^2+\frac{1}{4}\omega^2\right)$\\
{\bf Proca}&&\\
$c_1=m^2$&$c_1=m^2$&$c_1=m^2$\\
$c_2=2 m^2$ & $c_2=2 m^2$ & $c_2=2 m^2$
\end{tabular}
\end{center} 
We specify that we cannot force the fermions or bosons to satisfy other general rules in non-minimal  couplings \cite{BD} since then the coupling parameters might depend on  spin which is unacceptable. 

The conclusion is that on the hyperbolic spacetimes, dS and CAdS, one cannot establish an universal mass condition. Nevertheless, we may accept two separate mass conditions, $M=m$  for fermions and $c_1=m^2$ for bosons, but we believe that now it is premature to draw definitive conclusion based only on the above few examples. 

\section{Concluding remarks}

We demonstrated the CR-UIR equivalence in the case of the free fields defined on the CAdS spacetime, showing how the principal invariants depend on the effective mass and spin. 

We used the algebraic method of the ladder operators that may be applied here since we succeeded to find how the angular momentum operators act on the ground state. All these results were obtained by using algebraic codes on computer that work very well in Cartesian coordinates and Cartesian gauge.

However, the advantages of this algebraic method stop here since this is not suitable for studying the orthonormalization of the energy basis because of the interference among the Heisenberg-type algebras produced by of the last term of Eq. (\ref{AAHJ}).  For this reason, the algebraic approach must be combined with analytical methods, in charts with spherical coordinates, for deriving orthonormalized quantum modes as eigenfunctions of several sets of commuting operators, as in the cases of the Klein-Gordon \cite{AdS4,Cq1} or Dirac \cite{Cq2,Cq3} fields. 

The results presented here complete our image about the covariant fields on maximally symmetric specetimes, helping us to understand the influence of gravity on the basic properties of these fields.

\appendix
\section{Finite-dimensional  reps. of the $sl(2,{\Bbb C})$ algebra}

The standard basis of the $sl(2,{\Bbb C})$ algebra is formed by the generators 
$J_i$ and $K_i$ that satisfy \cite{WKT}
\begin{equation}
[J_i,J_j]=i\varepsilon_{ijk}J_k\,,\quad [J_i,K_j]=i\varepsilon_{ijk}K_k\,, \quad
[K_i,K_j]=-i\varepsilon_{ijk}J_k\,,
\end{equation} 
having the Casimir operators $c_1=i{J}_i {K}_i$ and $c_2={J}^2-{ K}^2$. The linear combinations $A_i=\frac{1}{2}\left(J_i +i K_i\right)$ and $B_i=\frac{1}{2}\left(J_i-i K_i\right)$ form two independent $su(2)$ algebras satisfying
\begin{equation}
[A_i,A_j]=i\varepsilon_{ijk}A_k\,,\quad [B_i,B_j]=i\varepsilon_{ijk}B_k\,, \quad
[A_i,B_j]=0\,.
\end{equation} 
Consequently, any finite-dimensional irrep.  $\tau=(j_1, j_2)$ is carried by the space ${\cal V}_{\tau}={\cal V}_{j_1}\otimes {\cal V}_{j_2}$ of the direct product $(j_1)\otimes (j_2)$ of the UIRs $(j_1)$  and $(j_2)$ of the $su(2)$ algebras $(A_i)$ and respectively $(B_i)$. These irreps. are labeled either by the $su(2)$ weights $(j_1,j_2)$ or giving the values of the Casimir operators $c_1=j_1(j_1+1)-j_2(j_2+1)$ and $c_2=2[j_1(j_1+1)+j_2(j_2+1)]$.

The fundamental reps. defining the $sl(2,{\Bbb C})$ algebra are either the irrep. 
$(\frac{1}{2},0)$  or the irrep. $(0, \frac{1}{2})$ whose  direct sum form the spinor  rep. $\rho_D= (\frac{1}{2},0)\oplus (0, \frac{1}{2})$ of the Dirac theory. This conjecture can be generalized easily considering pairs of {\em adjoint} irreps.,   $\tau=(j_1,j_2)$ and  $\dot \tau=(j_2,j_1)$, which have the same spin content while  their generators are related as ${J}_i^{(\dot\tau)}={J}_i^{(\tau)}$ and  ${K}_i^{(\dot\tau)}=-{K}_i^{(\tau)}$. Since  the operators ${A}_i$ and ${B}_i$ are Hermitian, generating  UIRs of the $su(2)$ algebra,  we must have ${J}_i^+={J}_i$ and  ${K}_i^+=-{K}_i$  such that we can write
\begin{equation}
({J}_i^{(\tau)})^+={J}_i^{(\tau)}\,, \quad ({K}_i^{(\tau)})^+={K}_i^{(\dot\tau)}\,.
\end{equation} 
Hereby we conclude that the invariant forms can be constructed only when we use {\em symmetric} reps. $\rho=\cdots \tau_1\oplus\tau_2\cdots \dot \tau_1\oplus\dot\tau_2\cdots$
containing only pairs of adjoint irreps. and/or self-adjoint irreps. $\tau=\dot\tau=(j,j)$. 
Then, the matrix $\gamma_{(\rho)}$ may be constructed having the matrix elements 
\begin{equation}
\langle \tau_1,s_1\sigma_1|\gamma_{(\rho)}|\tau_2,s_2\sigma_2\rangle=\delta_{\tau_1\dot\tau_2}\delta_{s_1s_2}\delta_{\sigma_1\sigma_2}\,,
\end{equation}
in the canonical basis $\{|\tau j\lambda\rangle\,| \tau \in \rho\}$ \cite{WKT}. Moreover, for such reps. we can construct at any time the charge conjugation matrix ${C}_{(\rho)}$ having the matrix elements \cite{CdSnew}
\begin{equation}
\langle \tau_1,s_1\sigma_1|C_{(\rho)}|\tau_2,s_2\sigma_2\rangle=\eta(\tau_1)\delta_{\tau_1\dot\tau_2}\delta_{s_1s_2}(-1)^{s_1-\sigma_1}\delta_{\sigma_1,-\sigma_2}\,,
\end{equation}
with $\eta(\tau)=\pm 1$. We note that the canonical basis defines the {\em chiral} rep. while a new basis in which $\gamma_{(\rho)}$ becomes diagonal gives the so called standard rep.. This terminology comes from the Dirac theory where $\gamma_{(\rho_D)}=\gamma^0$.

\subsection*{Acknowledgements}

This work was supported by a grant of the Ministry of National Education and Scientific Research, RDI Programme for Space Technology and Advanced Research - STAR, project number 181/20.07.2017.

\end{document}